\documentclass[twocolumn,pre,floatfix]{revtex4}
\usepackage{psfrag,epsfig,amsfonts,amssymb,amsmath,wasysym,bm}
\usepackage{dcolumn}
\usepackage{bbold}
\usepackage[normalem]{ulem}
\usepackage{color}
\usepackage{tabularx}

\usepackage{enumitem} 

\usepackage{hyperref}

\newcommand{\ket}[1]{\lvert #1 \rangle} 	
\newcommand{\bra}[1]{\langle #1 \rvert}	
\newcommand{\ketV}[1]{\lvert #1 \rangle_{\! V}} 	
\newcommand{\braV}[1]{{_V\!}\langle #1 \rvert}	
\newcommand{\mc}{\mathcal}	

\newcommand{\<}{\left\langle}	
\renewcommand{\>}{\right\rangle}	

\newcommand{\e}{\mathrm{e}}		
\newcommand{\I}{\mathrm{i}}		

\newcommand{\rhoMC}{\rho_{\mathrm{mc}}}
\newcommand{\rhoeq}{\rho_{\mathrm{eq}}}
\newcommand{\rhoForward}{\rho_{\mathrm{f}}}
\newcommand{\rhoScrambling}{\rho_{\mathrm{s}}}
\newcommand{\rhoBackward}{\rho_{\mathrm{b}}}

\newcommand{\rhoTarget}{\rho_{\mathrm{T}}}
\newcommand{\rhoTargetPert}{\rho^\prime_{\mathrm{T}}}
\newcommand{\rhoRev}{\rho_{\mathrm{R}}}
\newcommand{\rhoRevPert}{\rho^\prime_{\mathrm{R}}}

\newcommand{\dosV}{D_V}
\newcommand{\dosVft}{\hat{D}_V}

\newcommand{\dA}{\mathcal{A}}

\newcommand{\tr}{\mbox{Tr}}

\newcommand{\da}{\Delta_{\! A}}
\newcommand{\dat}{\Delta_{\! A(t)}}

\providecommand{\opnorm}[1]{\|#1\|}

\providecommand{\avv}[1]{\mathbb{E}\!\left[#1\right]}
\providecommand{\avu}[1]{\mathbb{E}[#1]}
\providecommand{\avvu}[1]{\mathbb{E}\!\left[#1\right]}

\renewcommand{\d}{\mathrm{d}} 
\newcommand{\lmat}{\left( \begin{matrix}}	
\newcommand{\rmat}{\end{matrix} \right)}	

\begin{document}

\title{Persistent many-body quantum echoes}
\author{Lennart Dabelow}
\author{Peter Reimann}
\affiliation{Fakult\"at f\"ur Physik, 
Universit\"at Bielefeld, 
33615 Bielefeld, Germany}
\date{\today}

\begin{abstract}
We consider quantum many-body 
systems evolving under a time-independent Hamiltonian 
$H$ from a nonequilibrium initial state at time $t=0$
towards a close-to-equilibrium state at time $t=\tau$.
Subsequently, this state is slightly perturbed and
finally propagated for another time period $\tau$
under the inverted Hamiltonian $-H$.
The entire procedure may also be viewed as
an imperfect time inversion or ``echo dynamics''.
We unravel a remarkable persistence of such
dynamics with respect to the observable
deviations of the time-dependent expectation 
values from 
the equilibrium expectation value:
For most perturbations, the deviations in the 
final state are essentially independent of the 
inversion time point $\tau$.
Our quantitative analytical predictions
compare very well with exact 
numerical results.
\end{abstract}

\maketitle

\section{Introduction}
\label{s1}
A trademark of {\em classical} chaos 
\cite{ott94, lic92, lak03}
is the
sensitivity against small perturbations:
Tiny changes in the initial conditions 
grow exponentially in time. 
While this effect is 
readily observable
in low-dimensional systems, things are less 
obvious in the case of
many-body
dynamics.
Indeed, already the mere fact that macroscopic 
experiments are reproducible suggests that
the practically unavoidable small 
differences in the microscopic initial 
conditions usually do {\em not} result 
in rapidly growing differences of 
the actually measurable quantities.

A first key role in this context is played by the
following purely
geometric effect:
Given a system state (classical phase space point) 
$\phi$ and an observable (phase space function) 
$A(\phi)$, small variations $\delta\phi$
are best detectable if they point into 
the direction $n(\phi)$ of the (normalized) 
gradient of $A(\phi)$.
In contrast, variations perpendicular
to $n(\phi)$ are (in leading order) undetectable
by means of this observable.
Since the subspace orthogonal to
$n(\phi)$ is of very high dimensionality,
the vast majority of all possible variations 
$\delta\phi$ exhibit very small
components parallel to $n(\phi)$, 
and hence the observable changes 
$A(\phi+\delta\phi)-A(\phi)$ 
become unresolvably small.

Focusing on isolated 
systems
with a large (but finite) number of degrees of freedom,
yet another key role
is played by the fact that they (generically)
exhibit equilibration 
\cite{dal16,gog16,mor18,rei13}, 
i.e., most initial states $\phi(0)$ 
evolve in time so that the observable 
$A(\phi(t))$ initially shows a transient 
relaxation behavior, and later 
remains very close to a constant value 
(apart from exceedingly rare 
recurrences).
Moreover, different initial
values $A(\phi(0))$ will result 
in practically identical long-time
averages (equilibrium values) of
$A(\phi(t))$ under quite weak 
requirements, namely similar values 
of the energy and of possibly 
existing further conserved 
quantities.
In other words, initial differences
are largely ``forgotten'' at later times.

In short, it would be nice to directly
``see'' (by means of an observable)
the exponential growth of small initial 
perturbations in isolated many-body
systems, but 
this seems not
easy to achieve in practice.

A first step 
consists in evolving $\phi(t)$ up to some
time point $\tau$, comparable to or larger
than the equilibration time of $A(\phi(t))$, 
and considering
the so-obtained state $\phi(\tau)$ 
as a ``new'' initial condition, which 
is then evolved under the time-inverted 
dynamics (so-called echo dynamics 
\cite{hah50}).
After another time span $\tau$ one thus
recovers the original initial state 
$\phi(0)$.
However, 
intuitively it now seems reasonable
to expect that this inverted evolution will
exhibit a more easily observable instability 
in the sense that small perturbations of 
$\phi(\tau)$ may typically lead to 
significant deviations
of $A(\phi(2\tau))$
from the unperturbed ``final'' value 
$A(\phi(0))$ of the echo dynamics. 
Furthermore, one may expect that this 
instability is closely related to 
the sensitivity of the original
(forward-in-time) solution $\phi(t)$
against small initial perturbations.
Though all those expectations 
appear intuitively quite plausible,
providing a more tangible justification 
seems to be a surprisingly difficult task.
Fortunately, such a more detailed analytical 
and numerical confirmation
has been provided, for instance,
in Refs. \cite{wij12, wij13}.

A second step in making this exponential
instability directly observable is based on
the fact that the time inversion is
equivalent to a sign change of the 
Hamiltonian, which in turn can be
experimentally realized -- at least for spin 
systems -- by inverting
the relative orientation of the externally 
applied (and possibly inhomogeneous)
magnetic field in cases with non-interacting spins 
(so-called Hahn echo \cite{hah50}),
and by means of more sophisticated
``tricks'' in cases with interacting 
spins \cite{sch69, rhi70, rhi71, zha92, kim92, haf96, lev98, usa98}.

The main topic of our present work is the
corresponding echo dynamics within a 
{\em quantum mechanical} instead of a
classical approach.
Besides its conceptual interest,
this is obviously also of relevance with 
respect to the 
experimental realizations in terms of spin systems, 
cold atoms, and quantum simulators \cite{hah50,rhi70,rhi71,zha92,kim92, haf96, lev98,usa98, wid08, lin16, gar17, nik20}.

The above arguments of why small initial 
perturbations usually also remain unobservable at all later times
(or, equivalently, why macroscopic 
experiments are reproducible) can be 
readily adapted to the quantum case.
But what about the echo dynamics?

In related previous works, the main 
focus was on the similarities and 
differences between classical
and quantum spin models \cite{fin14,els15} and
on questions of
whether and in which sense
the aforementioned signatures of classical chaos
can indeed be recovered also in quantum systems 
in the thermodynamic 
limit \cite{sch16,sch18,sch19}.
Notably, the thermodynamic
limit was thus formally carried out 
before letting the echo imperfections 
go to zero,
and it was found that
the two limits actually do 
not commute \cite{sch19}.

In our present approach, the general setup
is similar as in those previous works 
\cite{fin14,els15,sch16,sch18,sch19},
whereas the main focus is laid on a somewhat 
different issue, namely on quantitative 
analytical approximations for generic, 
large but finite systems in the 
presence of sufficiently small 
(but finite) imperfections.
The pertinent key results of our paper
consist in the prediction~\eqref{eq:IIC:TypEcho}
about the detailed behavior of the observable 
echo dynamics, and in the demonstration that 
it agrees very well with numerical results.
In particular, the analytical result~\eqref{eq:IIC:TypEcho} predicts that the observable perturbation of the quantum echo due to tiny imperfections during the reversal procedure does not grow \emph{at all} upon increasing the inversion time point $\tau$, and this behavior is indeed recovered numerically up to possible transient effects for very short values of $\tau$.

The paper is organized as follows:
In Sec.~\ref{s3}, we introduce the general setting of the considered echo dynamics and types of imperfections.
The main result is then derived and verified numerically in Sec.~\ref{s4}, before we combine it with results from~\cite{dab20} for a related scenario in Sec.~\ref{s5}.
Finally, we summarize and discuss our findings in Sec.~\ref{s6}.

\section{Echoes and imperfections}
\label{s3}
In this section, we introduce the general
framework of our echo approach and the objectives of the later
sections.

To begin with, we consider a
finite, isolated many-body quantum system 
with time-independent Hamiltonian
\begin{equation}
\label{eq:H}
	H = \sum_n E_n \, \ket{n} \bra{n} \,.
\end{equation}
Before turning to questions of time reversal 
and its instability against small perturbations,
we collect the essential properties 
of the ``clean'' (time-independent)
dynamics.

If the system is prepared in a (pure or mixed) 
state with density operator $\rho(0)$ at time $t = 0$,
it then evolves according to the Schr\"odinger 
or von Neumann equation, so that the state at 
any later time $t>0$ is 
$\rho(t) := \e^{-\I H t} \rho(0) \e^{\I H t}$ 
($\hbar = 1$).
Given some observable (self-adjoint operator) $A$,
its expectation value in an arbitrary state 
$\rho$ 
follows as
$\langle A \rangle_{\!\rho} := \tr\{ \rho A \}$.
For the time evolution of the 
expectation value in the system 
state $\rho(t)$ we thus obtain
\begin{equation}
\label{eq:TimeEvo:A}
	\< A \>_{\!\rho(t)} 
= 
\sum_{m, n} \e^{\I (E_n - E_m) t} 
\, 
\bra{m} \rho(0) \ket{n} \, \bra{n} A \ket{m} 
\,.
\end{equation}
Despite the quasiperiodic nature of this expectation 
value, a many-body system is known to exhibit
equilibration under rather weak conditions 
\cite{rei08, lin09, sho12}, i.e.,
after the initial transients have died out, 
it spends most of its time close to the  
time-independent value
$\< A \>_{\!\rhoeq}$,
where
$\rhoeq$
is (essentially) the long-time average 
of $\rho(t)$ \cite{rei12}.
Moreover, generic (nonintegrable) systems also thermalize \cite{dal16,gog16,mor18},
meaning that $\< A \>_{\!\rhoeq}$ can be identified
with $\< A \>_{\!\rhoMC}$, where $\rhoMC$ is
the pertinent microcanonical ensemble for
the isolated system at hand.

We emphasize that our main result
will actually turn out to apply
irrespective of whether or not the system 
exhibits equilibration, let alone thermalization.
However, if it does thermalize, 
then to
study the effect of
imperfect echo dynamics,
the system must start out sufficiently far from 
equilibrium, i.e., there must exist an initial
time interval during which 
\begin{equation}
\label{eq:DevAFromEq}
	\dA(t) := \<A\>_{\!\rho(t)}  - \< A \>_{\!\rhoMC} 
\end{equation}
notably deviates from zero.
Assuming that the system is in a nonequilibrium 
state at time $t = 0$, we then probe how special 
this situation is by exploring how difficult it is
to return to this 
state by an effective, but possibly imperfect 
reversal of the dynamics after the system
has
relaxed for some time.
In other words, imagine that the many-body system 
with Hamiltonian $H$ is prepared in a well-controlled 
far-from-equilibrium state $\rho(0) = \rhoTarget$, 
called {\em target state}, such that $\dA(0) \neq 0$.
Then, the system evolves according to the Hamiltonian 
$H$ until time $\tau$, resulting in a {\em return state}
$\rhoRev:=\rho(\tau)$.
From Eq.~\eqref{eq:TimeEvo:A}, we understand 
that evolving the system backward in time 
with the Hamiltonian $H$ is equivalent to evolving it
forward in time with the ``backward'' Hamiltonian 
$-H$.
Switching to this backward Hamiltonian at $t = \tau$, 
the resulting evolution from $\rhoRev$ to 
$\rhoTarget$ during the interval $[\tau, 2\tau]$ 
is just the reflection of the initial process 
during $[0, \tau]$ and constitutes our reference 
dynamics, i.e., the perfect time reversal with
\begin{equation}
\label{eq:PTR}
	\dA(\tau + t) = \dA(\tau - t)
\end{equation}
for any $t \in [0, \tau]$.

Note that even in cases which are not expected to exhibit
thermalization,
the difference in (\ref{eq:DevAFromEq}) will usually be quite
appreciable at early times $t$, but may possibly no longer approach 
zero for (most) sufficiently late times.
In any case, the quantity in (\ref{eq:DevAFromEq})
will later turn out to play a crucial role within 
the analytical approach adopted in our present paper
(see Sec. \ref{s4b} below).

Finally, two different types of imperfections 
are often considered \cite{fin14,els15,sch16,sch18,sch19} 
leading to deviations from the above 
reference scenario.
In the first case [sketched in 
Eq.~\eqref{eq:IIC:Protocol} below], 
one slightly perturbs the return state 
$\rhoRev$ by evolving it for a short 
time $\delta$ with a
``scrambling Hamiltonian''
$V$ \cite{f1} to obtain a 
different state $\rhoRevPert$, modeling 
an imprecise preparation of initial 
conditions for the time-reversed setup.
Thereafter, the system is evolved with 
the ``true'' backward Hamiltonian $-H$.
(Strictly speaking, the backward evolution 
thus takes place in the time interval 
$[\tau+\delta,2\tau+\delta]$, but for 
simplicity we will sometimes
approximately denote it as $[\tau,2\tau]$.)
In the second case [cf.\ Eq.~\eqref{eq:ITR:Protocol}], 
one starts from the original return state $\rhoRev$, 
but now includes a small perturbation in the 
backward Hamiltonian $H' := -H + \epsilon W$, 
modeling an imperfect implementation of the 
time-reversed system.

Henceforth, those two types of perturbations 
will be referred to as {\em imperfect preparation}
and {\em imperfect reversal}, respectively.

The main focus of our present paper
will be on perturbations of the first 
type (imperfect preparation).
But for the sake of completeness and 
comparability, we will also briefly 
summarize our previous findings 
from Ref.~\cite{dab20} for perturbations 
of the second type.
Moreover, we will provide the pertinent
extension when both types of perturbations
are simultaneously present.

In either case, one intuitively expects that a
generic imperfection of the time-reversed dynamics 
typically brings the system closer to equilibrium 
because it impairs the fine-tuned correlations 
between state and observable that are vital to obtain
a reasonably far-from-equilibrium expectation value
in (\ref{eq:TimeEvo:A}).
Hence, the imperfect backward dynamics will usually 
remain closer to the equilibrium state than the 
forward one, i.e., 
\begin{equation}
\label{eq:ITR}
	\lvert \dA(\tau + t) \rvert \lesssim \lvert \dA(\tau - t) \rvert \,,
	\qquad t \in [0, \tau] \,.
\end{equation}
The chaoticity and irreversibility of the many-body
dynamics may thus be gauged in terms of the sensitivity
of the deviations between the perfect and scrambled 
echo dynamics with respect to the two above introduced scrambling 
parameters $\delta$ and $\epsilon$:
The more $\dA(\tau+t)$ is attenuated with 
increasing $\delta$ 
or $\epsilon$ compared to $\dA(\tau - t)$, the 
harder it is to design a reversible process 
and the more extraordinary or special are the 
initially probed nonequilibrium states.

Accordingly, the ratio $\dA(\tau + t) / \dA(\tau - t)$ 
for times $t \in [0, \tau]$
is at the heart of our present study,
most importantly in 
the region around the revival or {\em echo peak}
of $\dA(\tau + t)$ at $t \approx \tau$, 
where observable deviations from 
equilibrium will
stand out most prominently,
cf.~(\ref{eq:DevAFromEq}).

Our main results will be analytical predictions 
for the ratio $\dA(\tau + t) / \dA(\tau - t)$, 
based on a few basic properties of 
the scrambling operators $V$ and $W$.
In the derivation, we assume that these operators are
-- essentially by definition -- 
uncontrolled (random) 
and, in particular, independent 
of the actual system Hamiltonian $H$, 
in a sense that will become 
more clear below.

Finally, we will also verify our predictions 
by comparison with numerically exact results
for a spin-$\frac{1}{2}$ XXX model.

\section{Imperfect preparation}
\label{s4}
\subsection{Setup}
\label{s4a}
As outlined above, our echo protocol modeling
an imperfect initial state for the backward 
evolution consists of three phases:
After preparing the system in the target state 
$\rhoTarget$, it is first evolved with the forward 
Hamiltonian $H$ for a time $\tau$ 
to reach the return state $\rhoRev$.
From there, second, the system is subject to a 
perturbing Hamiltonian $V$ for the short 
scrambling time $\delta$, yielding the state 
$\rhoRevPert$.
Third, we then evolve it backwards with the 
Hamiltonian $-H$ for a time $\tau$, ending 
up in a state $\rhoTargetPert$.
The entire process can thus be summarized as
\begin{equation}
\label{eq:IIC:Protocol}
	\rhoTarget \xrightarrow[H]{\;\;\;\;\tau\;\;\;\;} \rhoRev 
	\xrightarrow[V]{\;\;\;\;\delta\;\;\;\;} \rhoRevPert \xrightarrow[-H]{\;\;\;\;\tau\;\;\;\;} \rhoTargetPert \,.
\end{equation}
By analogy with classical systems and their pertinent indicators of chaos \cite{ott94, lic92},
one may roughly regard this 
approach as a probe for sensitive dependence 
on initial conditions since it quantifies how 
a small difference between the states 
$\rhoRev$ and $\rhoRevPert$ translates into 
a difference in the states $\rhoTarget$ and 
$\rhoTargetPert$ after time $\tau$. 
In this sense, the scrambling 
phase taking $\rhoRev$ to $\rhoRevPert$ need not necessarily be viewed as an actual time evolution.
Instead, more abstractly, one may assume a Lie group perspective and regard $\e^{\I V \delta}$ as a ``rotation'' in Hilbert space by an angle $\delta$ in the direction $V$.
When comparing different systems, the operator $V$ should then be properly normalized.
Nevertheless, for notational convenience, we will include 
the scrambling 
as part of the time evolution in the following.

We denote the time-dependent state of the system 
in the forward, scrambling, 
and backward phases by
\begin{subequations}
\label{eq:IIC:Phases}
\begin{alignat}{2}
\label{eq:IIC:Phases:Forward}
	\rhoForward(t) &:=& \,\,\e^{-\I H t} &\,\rhoTarget\, \e^{\I H t} \,, \\
\label{eq:IIC:Phases:Scrambling}
	\rhoScrambling(t) &:=& \,\,\e^{-\I V t} &\,\rhoRev\, \e^{\I V t} \,, \\
\label{eq:IIC:Phases:Backward}
	\rhoBackward(t) &:=& \,\,\e^{\I H t} &\,\rhoRevPert\, \e^{-\I H t} \,,
\end{alignat}
\end{subequations}
respectively.
We also write $\rho(t)$ to refer to the state 
during the entire process, i.e., 
$\rho(t) := \rhoForward(t)$ for 
$t \in [0, \tau]$, $\rho(t) := \rhoScrambling(t - \tau)$ 
for $t \in [\tau, \tau + \delta]$, 
and $\rho(t) := \rhoBackward(t - \tau - \delta)$ 
for $t \in [\tau + \delta, 2\tau + \delta]$.

An implicit assumption in all what follows is 
that the considered many-body system under study 
is finite and exhibits a well-defined macroscopic 
energy $E$.
In particular, it is understood that the perturbation 
taking the return state $\rhoRev$ to $\rhoRevPert$ 
is so small that it does not change this energy, i.e., 
there should be no macroscopically measurable energy 
intake during the scrambling 
phase. 
Consequently, the state $\rho(t)$ at any time can 
only significantly populate energy levels within a 
macroscopically small window $[E - \Delta E, E]$.
Therefore, we can and will implicitly restrict 
summations over energy levels to this window 
in the following.
Given the extremely small level spacing in 
many-body systems, the number of levels $N$ within 
such an energy shell is still exponentially large 
in the system's degrees of freedom $f$ \cite{lan70},
\begin{equation}
\label{eq:IIC:LevelsInShell}
	N = 10^{\mathcal{O}(f)} \gg 1 \,.
\end{equation}
Another temporary assumption we make is that the return state $\rhoRev$ populates the levels of the 
scrambling 
Hamiltonian $V$ approximately uniformly 
within the relevant energy interval $[E-\Delta E,E]$.
This assumption seems reasonable in the absence of 
any additional knowledge about the experimental 
imperfections modeled by $V$.
Should such information be available, it
is still possible to generalize our present approach 
along similar lines as 
in Ref.~\cite{rei19a} in order to account 
for significant non-uniformities in the level 
populations.
We will comment on the necessary modifications at the end of Sec.~\ref{s4d}.

Aiming at a prediction for the echo dynamics, 
we now turn to the third phase in the 
evolution~\eqref{eq:IIC:Protocol},
where the state of the system is 
[cf.~Eqs.~\eqref{eq:IIC:Phases}]
\begin{equation}
	\rhoBackward(t) = \e^{\I H t} \e^{-\I V \delta} \e^{-\I H \tau} \rhoTarget \e^{\I H \tau} \e^{\I V \delta} \e^{-\I H t} \,.
\label{13}
\end{equation}
Recalling that the eigenvalues and eigenstates 
of the Hamiltonian $H$ are $E_n$ and $\ket{n}$, 
cf.~Eq.~\eqref{eq:H}, and denoting those of $V$ 
by $E^V_\nu$ and $\ketV{\nu}$, respectively, the 
matrix elements $\bra{m} \rhoBackward(t) \ket{n}$ 
can be written as
\begin{equation}
\label{eq:IIC:TimeEvoBack:rho}
\begin{aligned}
	\bra{m} \rhoBackward(t) \ket{n}
		&= \sum_{k,l,\mu,\nu} \e^{-\I (E_n-E_m) t} \e^{\I (E^V_\nu-E^V_\mu) \delta} \e^{\I (E_l-E_k) \tau} \\
		&\qquad\qquad \times \bra{k} \rhoTarget \ket{l} \, U_{\mu k} U^*_{\mu m} U^*_{\nu l} U_{\nu n} \,.
\end{aligned}
\end{equation}
Here $U_{\mu k} := \braV{\mu} k \rangle$ are the 
transformation matrices between the eigenbases of 
$H$ and $V$.
For a particular observable $A$, the time-dependent 
expectation value is then given by
\begin{equation}
\label{eq:IIC:TimeEvoBack:A}
	\< A \>_{\!\rhoBackward(t)}
		= \sum_{m, n} \bra{m} \rhoBackward(t) \ket{n} \, \bra{n} A \ket{m} \,.
\end{equation}

We recall that the Hamiltonian $H$ corresponds to 
a {\em given} many-body quantum system, whereas the 
perturbation $V$ is supposed to describe an 
{\em uncontrolled} experimental uncertainty
in the time-reversal procedure with the
result of an 
imperfectly prepared return state $\rhoRevPert$.
In this spirit, we will thus model the scrambling 
Hamiltonian $V$ by a random Hermitian matrix, 
conforming with our 
partial knowledge and partial ignorance about the 
actual
experimental imperfections.

Regarding the statistical properties of $V$, 
we adopt the rather general framework from 
Ref.~\cite{rei16}, 
where eigenvalues and eigenvectors are 
considered as statistically independent quantities.
While the former (i.e., the $E^V_\nu$'s in 
(\ref{eq:IIC:TimeEvoBack:rho}))
are considered as ``given'', i.e.,
as largely arbitrary but fixed 
(non-random), the latter are generated according 
to a uniform (unbiased) distribution, i.e., the 
$U_{\mu k}$'s in (\ref{eq:IIC:TimeEvoBack:rho})
are sampled via the Haar measure of 
the appropriate symmetry group, e.g.\ the 
circular unitary ensemble (CUE) for 
$\mathrm{U}(N)$ or the circular orthogonal 
ensemble (COE) 
for symmetric unitary matrices.
For instance, the well-known standard case of 
a Gaussian unitary $V$-ensemble (GUE)
is recovered by combining the CUE with 
a semicircular density of states (DOS) 
for the eigenvalues \cite{haa10}.
In general, i.e., unless there is specific 
additional knowledge about the structure of 
typical experimental uncertainty in the 
time-reversal procedure, it may however be 
reasonable to also leave room for some different 
(non-semicircular) DOS, for instance
one which exhibits some similarity with
the DOS of the system Hamiltonian $H$.

This approach provides us with a useful prediction for the behavior of an actual system in two steps:
First, we calculate how such a perturbation affects the observed dynamics \emph{on average}.
Second, we show that the squared magnitude of fluctuations around the average is inversely proportional to the number of states $N$ in the pertinent energy shell.
For large $N$, this concentration of measure property ensures that a single realization will typically not deviate from the ensemble average in practice.
Bearing in mind Eq.~\eqref{eq:IIC:LevelsInShell}, 
we may then conclude that such a modeling should 
indeed account for most uncontrolled uncertainties 
in setting up the return state.

\subsection{Results}
\label{s4b}
The average effect of the scrambling 
with $\e^{-\I V \delta}$ 
in (\ref{13})
is obtained by averaging over the corresponding,
above specified $U$ ensemble in 
Eq.~\eqref{eq:IIC:TimeEvoBack:rho}.
Recalling that we keep the spectrum of 
$V$ fixed, the only random quantities 
are the four factors of $U_{\mu k}$, 
i.e., the eigenvectors of $V$ in the 
eigenbasis of $H$.
Denoting this average over the Haar-distributed 
$U_{\mu k}$ by $\avu{ \,\cdots }$,
we can consult Ref.~\cite{bro96} to 
find the general form
\begin{equation}
\label{eq:IIC:U4}
\begin{aligned}
	&\avu{ U_{\mu_1 k_1} U_{\mu_2 k_2} U^*_{\mu_1 l_1} U^*_{\mu_2 l_2} } \\
		& \; = v_{1,1} \left( \delta_{k_1 l_1} \delta_{k_2 l_2} + \delta_{\mu_1 \mu_2} \delta_{k_1 l_2} \delta_{k_2 l_1} \right) \\
		& \quad + v_2 \left( \delta_{k_1 l_2} \delta_{k_2 l_1} + \delta_{\mu_1 \mu_2} \delta_{k_1 l_1} \delta_{k_2 l_2} \right) ,
\end{aligned}
\end{equation}
where the so-called symmetry factors 
$v_{1,1}$ and $v_2$ are given by 
$v_{1,1} = \frac{1}{N^2 - 1}$ and 
$v_2 = -\frac{1}{N (N^2 - 1)}$ for 
$\mathrm{U}(N)$ and the CUE, or by
$v_{1,1} = \frac{N+2}{N(N+1)(N+3)}$ and 
$v_2 = -\frac{1}{N(N+1)(N+3)}$ 
for the COE.
For our purposes, where $N \gg 1$ 
according to~(\ref{eq:IIC:LevelsInShell}), 
we can approximate $v_{1,1} \simeq \frac{1}{N^2}$ 
and $v_2 \simeq -\frac{1}{N^3}$ in both 
cases.
Exploiting all this in carrying out the average 
of Eq.~\eqref{eq:IIC:TimeEvoBack:rho}, one
finally obtains
\begin{equation}
\label{eq:IIC:AvgTimeEvo:rho}
\begin{aligned}
	& \avu{ \bra{m} \rhoBackward(t) \ket{n} } \\
	& \; = \e^{\I (E_n - E_m) (\tau - t)} \bra{m} \rhoTarget \ket{n} \left[ \frac{1}{N^2} \sum_{\mu,\nu} \e^{\I (E_\nu^V - E_\mu^V) \delta} - \frac{1}{N} \right] \\
	& \qquad + \frac{\delta_{mn}}{N} \left[ 1 - \frac{1}{N^2} \sum_{\mu,\nu} \e^{\I (E_\nu^V - E_\mu^V) \delta} \right] .
\end{aligned}
\end{equation}
In the second line, we retrieve the forward state
\begin{equation}
	\bra{m} \rhoForward(\tau - t) \ket{n}
		= \e^{\I (E_n - E_m) (\tau - t)} \bra{m} \rhoTarget \ket{n}
\end{equation}
from~\eqref{eq:IIC:Phases:Forward}.
In the third line, we can identify 
the matrix elements of the microcanonical 
density operator $\rhoMC$,
\begin{equation}
\label{eq:rhoMC}
	\bra{m} \rhoMC \ket{n} = \frac{\delta_{mn}}{N} \,.
\end{equation}
The remaining terms can be conveniently rewritten
by means of the DOS
of the scrambling 
Hamiltonian $V$ in the 
relevant energy shell,
\begin{equation}
\label{eq:DOS:V}
	\dosV(E) := \frac{1}{N} \sum_\nu \delta(E - E_\nu^V)
	\ ,
\end{equation} 
and its Fourier transform,
\begin{equation}
\label{eq:DOS:V:FT}
	\dosVft(t) := \int \d E \; \dosV(E) \, \e^{\I E t} \,.
\end{equation}
Altogether, Eq.~\eqref{eq:IIC:AvgTimeEvo:rho} then 
takes the form
\begin{equation}
\label{eq:IIC:AvgTimeEvo:rho2}
	\avu{ \rhoBackward(t) }
		= \rhoForward(\tau - t) \left[ \lvert \dosVft(\delta) \rvert^2 - \frac{1}{N} \right]
			+ \rhoMC \left[ 1 - \lvert \dosVft(\delta) \rvert^2 \right] .
\end{equation}
Observing that the $1/N$ term is negligible 
due to $N \gg 1$, 
and exploiting the definition~\eqref{eq:DevAFromEq} with $\rho(t)$ 
as introduced below~\eqref{eq:IIC:Phases}, we 
then find
\begin{equation}
\label{eq:IIC:AvgEcho}
	\frac{ \avu{ \dA(\tau + \delta + t) } }{ \dA(\tau - t) }
		= \lvert \dosVft(\delta) \rvert^2 
		\qquad
		(0 \leq t \leq \tau) \,.
\end{equation}
This is the first key result of this section, 
relating the relative strength of the ensemble-averaged
echo signal under imperfect preparation 
of the return state to the density of states 
of the scrambling 
operator.

To assess deviations between the observed dynamics for one 
particular realization of $V$ and the 
above derived average behavior,
we consider the variance of $\dA(\tau + \delta + t)$.
In view of~\eqref{eq:IIC:TimeEvoBack:rho}
 and~\eqref{eq:IIC:TimeEvoBack:A}, 
we thus need averages over products of eight 
matrix elements $U_{\mu k}$.
Employing Ref.~\cite{bro96} once again, the result is structurally 
similar to Eq.~\eqref{eq:IIC:U4}, i.e., it consists of Kronecker-$\delta$ 
contractions multiplied by symmetry factors such as $v_{1,1}$, etc.
Referring to Appendix~\ref{app2} for the details, 
one finally obtains
\begin{equation}
\label{eq:IIC:Variance}
	\avvu{ \dA(\tau \!+\! \delta \!+\! t)^2 } - \avvu{ \dA(\tau \!+\! \delta \!+\! t) }^{\,2}
		\leq  11\frac{\da^2}{N} + \mathcal{O}\!\left(\frac{1}{N^2}\right) ,
\end{equation}
where $\da$ indicates the measurement 
range of the observable $A$, i.e., the 
difference between the largest and
smallest possible measurement outcomes
(eigenvalues).
Similarly as below Eq.~\eqref{eq:IIC:U4}, the differences
between the CUE and the COE only appear in the
higher-order corrections (last term in (\ref{eq:IIC:Variance})),
and similarly as below (\ref{eq:IIC:AvgTimeEvo:rho2}), 
these corrections will be henceforth neglected.

The result~\eqref{eq:IIC:Variance} establishes the following 
concentration of measure
property of the considered ensembles of $V$ 
operators:
Roughly speaking, the fluctuations for one particular $V$ 
typically decrease like $1/\sqrt{N}$ and thus exponentially 
fast in the 
system's degrees of freedom, see Eq.~\eqref{eq:IIC:LevelsInShell}.
More precisely, Chebyshev's inequality implies that the 
probability for $\dA(\tau+\delta+t)$ to differ by more than 
$\da / N^{1/3}$ from the average $\avu{ \dA(\tau+\delta+t) }$ 
at a certain instant in time $t$ is less than $11/N^{1/3}$.
This is our second key result, providing
predictive power to our first key result from 
Eq.~\eqref{eq:IIC:AvgEcho}.
Namely, since observable deviations 
from the average are extremely
scarce for systems with a reasonably large number of degrees of freedom,
we can conclude that
\begin{equation}
\label{eq:IIC:TypEcho}
	\frac{ \dA(\tau + \delta + t) }{ \dA(\tau - t) }
		= \lvert \dosVft(\delta) \rvert^2
		\qquad
		(0 \leq t \leq \tau)
\end{equation}
amounts to a very good approximation for nearly all 
imperfections in the return state preparation which can be 
modeled by~\eqref{eq:IIC:Protocol}.

Eq.~(\ref{eq:IIC:TypEcho}) is our main result regarding the 
generic echo dynamics under imperfect initial conditions 
for the backward evolution.
As announced below (\ref{eq:PTR}), we see that the 
quantity from (\ref{eq:DevAFromEq}) indeed plays a crucial
role, independently of whether or not the considered system is
expected to exhibit thermalization.

The particularly interesting revival or echo peak signal
is recovered by choosing $t=\tau$ in Eq.~(\ref{eq:IIC:TypEcho}).
Moreover, it can be recast
by means of~(\ref{eq:IIC:TypEcho}) 
and~(\ref{eq:DevAFromEq}) into the equivalent form
\begin{equation}
\label{x5}
	\<A\>_{\!\rho(2\tau+\delta)} - \<A\>_{\!\rho(0)}
	= [ \< A \>_{\!\rhoMC} - \<A\>_{\!\rho(0)} ] \, [1- \lvert \dosVft(\delta) \rvert^2] \,.
\end{equation}
Noting that $\lvert \dosVft(\delta)\rvert^2 \simeq \lvert \tr[\rhoRev \e^{-\I V \delta}] \rvert^2$ under the assumption [cf.~below Eq.~\eqref{eq:IIC:LevelsInShell}] of an approximately uniformly spread-out return state $\rhoRev$ in the eigenbasis of $V$ \cite{tor14},
this result also confirms the finite-size analysis of Ref.~\cite{sch19} within the present typicality approach.

According to~\eqref{eq:IIC:TypEcho}, we thus expect that the 
``intensity'' of the revival dynamics 
(left-hand side)
is determined by the Fourier transform 
$\dosVft(\delta)$ of the perturbation's density 
of states evaluated at the scrambling 
time $\delta$.
The longer this scrambling 
time, the farther the effective return state is 
carried away from the perfect state $\rhoRev$, and the weaker the 
echo peak signal will typically be. 
Notably, though, 
the echo peak signal is predicted
to be independent of
the propagation time $\tau$, see also~(\ref{x5}).
In this respect,
the quantum echo dynamics is thus of a surprisingly
persistent character.
This should be contrasted with the classical case where -- as one might intuitively expect from chaos theory -- imperfectly prepared echo signals decay with $\tau$ \cite{wij12, wij13}.

\subsection{Example}
\label{s4c}
%
\begin{figure*}
\includegraphics[scale=1]{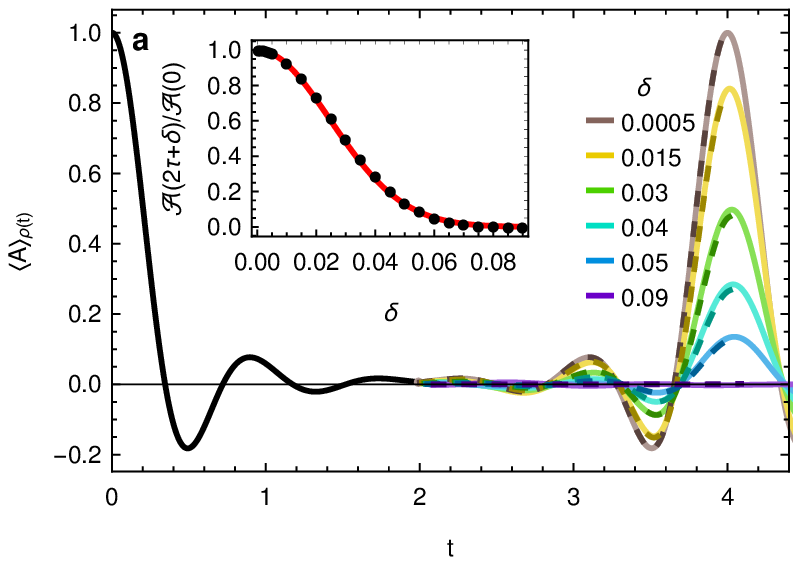}
\includegraphics[scale=1]{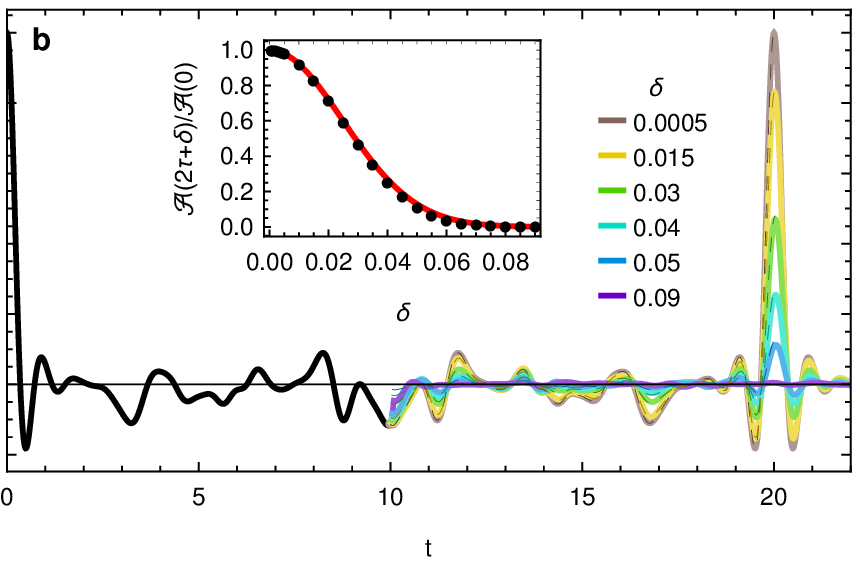}
\caption{
Time-dependent expectation values of the 
staggered magnetization~(\ref{eq:IIC:Example:A})
for the spin-$\frac{1}{2}$ XXX chain~(\ref{eq:IIC:Example:H})
with $L = 14$ under the ``imperfect preparation'' 
protocol~\eqref{eq:IIC:Protocol}, 
starting from a N\'{e}el target state $\rhoTarget$, 
for various scrambling 
times $\delta$.
The time reversal is initiated after time 
$\tau = 2$ in (a) and $\tau = 10$ in (b).
The perturbation Hamiltonian $V$ is of the 
``spin-glass'' form~\eqref{eq:IIC:Example:V}
for one realization of the normally distributed 
random couplings $J_{ij}^{\alpha\beta}$.
(For other realizations, we found 
practically the same results.)
Solid lines correspond to the numerical 
results using exact diagonalization,
color-coded as black 
for $t\in[0,\tau]$ (forward evolution), 
and brown-to-purple as indicated 
in the legend 
for $t\in[\tau,2\tau+\delta]$
with the peak height decreasing as $\delta$ increases.
Dashed lines show the analytical prediction for 
$t\in[\tau+\delta,2\tau+\delta]$
(backward or echo dynamics)
according to~\eqref{eq:IIC:TypEcho}, 
\eqref{x}, and \eqref{eq:IIC:Example:DHat},
complemented by (\ref{eq:DevAFromEq}) with 
$\< A \>_{\!\rhoMC}=0$ for symmetry reasons.
(The dashed lines are often hardly distinguishable 
from the corresponding solid lines.)
Insets: Ratio $\mc A(2\tau + \delta) / \mc A(0)$ of 
echo and initial peak heights  
as a function of the scrambling 
time $\delta$.
Data points are the numerical solutions, the solid red 
curve is the analytical prediction 
from~\eqref{eq:IIC:TypEcho}, \eqref{x}, and 
\eqref{eq:IIC:Example:DHat}.
}
\label{fig:IIC:SpinES}
\end{figure*}

As an illustrational example, we consider a spin-$\frac{1}{2}$ XXX 
chain of length $L$ with Hamiltonian
\begin{equation}
\label{eq:IIC:Example:H}
	H = - \sum_{i = 1}^{L-1} \bm\sigma_i \cdot \bm\sigma_{i+1} \,,
\end{equation}
where $\bm\sigma_i = (\sigma^x_i, \sigma^y_i, \sigma^z_i)$ is a vector 
collecting the Pauli matrices acting on site $i$.
The initial (target) state $\rhoTarget = \ket{\psi}\bra{\psi}$ is a N\'{e}el 
state, $\ket{\psi} = \ket{ \downarrow \uparrow \downarrow \uparrow \cdots }$,
and we observe the staggered magnetization
\begin{equation}
\label{eq:IIC:Example:A}
	A = \frac{1}{L} \sum_{i=1}^L (-1)^i \, \sigma_i^z \,.
\end{equation}
For the scrambling 
operator $V$, we choose
\begin{equation}
\label{eq:IIC:Example:V}
	V = \sum_{i<j} \sum_{\alpha, \beta = 1}^3 J_{ij}^{\alpha\beta} \sigma_i^\alpha \sigma_j^\beta \,,
\end{equation}
with independent, normally distributed 
couplings $J_{ij}^{\alpha\beta}$
of vanishing mean and unit variance.
Hence the direction in Hilbert space in which the return state $\rhoRev$ 
is perturbed is somewhat random and reflects the erratic character of 
the imperfections, but these still respect the overall setting and spin 
structure of the model under study.
Note that in the more abstract picture of $V$ as a generator of Hilbert space rotations [see below~\eqref{eq:IIC:Protocol}], $V$ should be normalized by dividing by the number of terms on the right-hand side in order to compare perturbation strengths across different system sizes.

For $L \gg 1$, the DOS of $V$ from~\eqref{eq:IIC:Example:V} is known to be a Gaussian 
with mean zero and variance \cite{erd14}
\begin{eqnarray}
\sigma_{\mathrm{DOS}}^2 = 9 L (L-1) / 2
\ .
\label{x}
\end{eqnarray}
With~\eqref{eq:DOS:V:FT} and~\eqref{eq:IIC:TypEcho}, we 
thus expect that the echo signal is characterized by
\begin{equation}
\label{eq:IIC:Example:DHat}
	\dosVft(\delta) = \e^{- \sigma_{\mathrm{DOS}}^2 \delta^2 / 2} \,.
\end{equation}

In Fig.~\ref{fig:IIC:SpinES} we show the resulting dynamics under the 
protocol~\eqref{eq:IIC:Protocol} for two durations $\tau$ and several 
scrambling 
times $\delta$,
and compare it to our analytical result~\eqref{eq:IIC:TypEcho}.
We emphasize that, by exploiting~\eqref{x} and 
\eqref{eq:IIC:Example:DHat}, this theoretical prediction 
does not involve any fit parameters.
Its agreement with the numerics
is practically 
flawless,  illustrating the decay of the echo peak as the 
intensity $\delta$ of the perturbation is increased.
At the same time, the persistence of the echo signal becomes apparent since the peak heights for fixed $\delta$ are identical for both values of $\tau$.
As we will discuss at the end of Sec.~\ref{s4d} (see also Appendix~\ref{app3}),
the numerically observed echo signal does show some $\tau$ dependence for very short $\tau \lesssim 1$ (below the relaxation time of the reference system),
so that deviations from our prediction occur in this regime.

\subsection{Perspectives}
\label{s4d}
It is instructive to reconsider the first of the three steps in 
(\ref{eq:IIC:Protocol})
by switching from the so-far adopted Schr\"odinger picture
to the equivalent Heisenberg picture and, moreover,
by considering time as evolving backward.
In other words, we observe that
$\rho(\tau-t)=U_{t}\rho(\tau)U^\dagger_{t}$ with 
backward-in-time propagator $U_t:=\e^{\I H t}$, 
and hence
\begin{eqnarray}
\langle A\rangle_{\!\rho(\tau-t)}
& = &
\tr\{\rho(\tau-t) A\}=\tr\{\rho(\tau) A(t)\}
\nonumber
\\
& = &
\tr\{\rhoRev A(t)\}
\ ,
\label{x1}
\\
A(t) & := & U^\dagger_{t}AU_{t}
\label{x2}
\end{eqnarray}
for all $t\in[0,\tau]$.
Likewise, upon observing that
$U_t$ from above amounts to the
forward-in-time propagator for the
inverted Hamiltonian $-H$,
the last step in (\ref{eq:IIC:Protocol}) 
can be rewritten as
\begin{eqnarray}
\langle A\rangle_{\!\rho(\tau+\delta+t)}
=\tr\{\rho(\tau+\delta) A(t)\}
=\tr\{\rhoRevPert A(t)\}
\label{x3}
\end{eqnarray}
for all $t\in[0,\tau]$.
Finally, one readily verifies that the 
measurement range $\dat$ 
[see below (\ref{eq:IIC:Variance})]
and the microcanonical
expectation value $\langle A(t)\rangle_{\!\rhoMC}$
are independent of $t$.
Altogether, one thus can conclude that 
if the relation (\ref{eq:IIC:TypEcho}) 
is valid for arbitrary observables 
at one single $t$ value, say $t=0$,
then it is valid for any given 
observable at arbitrary $t\in[0,\tau]$.

On the one hand, this 
conclusion is helpful to develop some intuitive 
understanding of
why the ratio on the left-hand side
of~(\ref{eq:IIC:TypEcho}) 
is predicted not to depend on $t$
(see right-hand side).
On the other hand,
in order to verify~(\ref{eq:IIC:TypEcho}) 
for arbitrary $A$ and $t$, it is sufficient
to solely focus on $t=0$.
Put differently, 
the first and the last of the 
three steps in (\ref{eq:IIC:Protocol}) are 
actually irrelevant for our present purpose
(both time evolutions can be 
omitted by considering 
the modified observables $\{A(t)\}_{t=0}^{\tau}$ 
instead of the original $A$).
As a consequence, the result 
(\ref{eq:IIC:TypEcho}) is expected to 
remain valid even for explicitly 
time-dependent Hamiltonians $H(t)$
in (\ref{eq:IIC:Protocol}).

Another immediate implication is that
the system does not need to (approximately)
approach a steady state at the end
of the first step in~(\ref{eq:IIC:Protocol})
(even if $\tau$ is very large),
i.e., neither equilibration nor 
thermalization are required.

\begin{figure}
\includegraphics[scale=1]{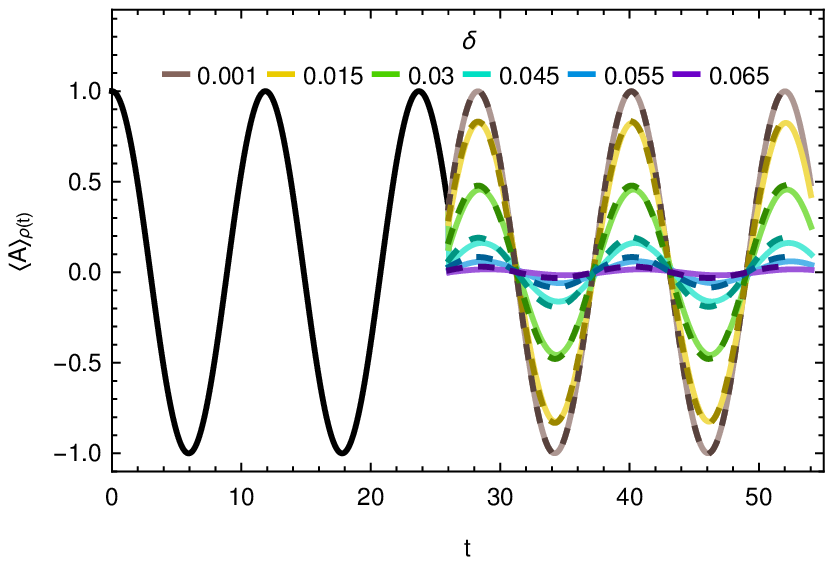}
\caption{
Same as in Fig.~\ref{fig:IIC:SpinES}, but for a different 
initial state $\rhoTarget$ and observable $A$, 
namely $\rhoTarget = \ket{\phi}\bra{\phi}$ 
with $\ket{\phi} = (\ket{n_1} + \ket{n_2})/\sqrt{2}$ 
and $A = (\ket{n_1} \bra{n_2} + \ket{n_2} \bra{n_1})/2$,
where $n_1 = 7936$ and $n_2 = 8448$
(hence $n_1+n_2=N=2^{14}$).
The time reversal is initiated after time $\tau = 26$.
The solid lines correspond to the numerical results, 
and the dashed lines to the analytical
prediction~\eqref{eq:IIC:TypEcho}, 
\eqref{x}, \eqref{eq:IIC:Example:DHat} 
with $\< A \>_{\!\rhoMC}=0$ due to symmetry.
(Similarly as in Fig. \ref{fig:IIC:SpinES},
dashed and solid lines are often hard to 
distinguish.)
}
\label{fig2}
\end{figure}

Fig.~\ref{fig2} depicts a particularly 
interesting example of this kind, which arises
when the initial state $\rho(0)$ only
populates two levels of the Hamiltonian $H$,
hence the expectation value in~(\ref{eq:TimeEvo:A})
exhibits perpetual oscillations in time.
According to~(\ref{eq:IIC:TypEcho}),
the echo dynamics thus also exhibits
everlasting oscillations in spite of the
scrambling 
effects via the second step 
in~(\ref{eq:IIC:Protocol}).
The reason is that also the level populations,
corresponding to observables of the form
$A=|n\rangle\langle n|$, are only moderately 
changed according to (\ref{eq:IIC:TypEcho})
for not too large scrambling 
time $\delta$.

Similarly as in Fig.~\ref{fig:IIC:SpinES},
our analytical theory again reproduces the 
numerically exact results in Fig. \ref{fig2} 
remarkably well without any free fit parameter.

Finally, we come back to the assumption from below Eq.~\eqref{eq:IIC:LevelsInShell}, stating that the return state $\rhoRev$ should occupy the energy levels of $V$ roughly uniformly.
If this is not the case, it is straightforward to adapt the formalism of Ref.~\cite{rei19a} to obtain a refined prediction for the echo dynamics that includes information about occupation imbalances.
The
essential modification affects the function 
$\dosVft(\delta)$ such that the integrand 
in~\eqref{eq:DOS:V:FT} receives an 
additional energy-dependent weight 
according to the populations of the levels 
around $E$.
While the definition of the function
$\dosVft(\delta)$ appearing on the right-hand 
side in (\ref{eq:IIC:TypEcho}) may thus 
be modified, the key point is that
it still remains independent of $t$.
Moreover, it also remains (approximately)
independent of $\tau$, provided $\tau$ 
exceeds the characteristic relaxation time scale
of the forward dynamics (governed by $H$ in~(\ref{eq:IIC:Protocol})).
For smaller $\tau$ values, things may
in general become considerably more complicated.
Indeed, for the same example as in 
Fig.~\ref{fig:IIC:SpinES}, we found that
the differences between numerics and 
theory quite notably increase in the 
regime $\tau \lesssim 1$, see also Appendix~\ref{app3}.
Analogous numerical findings for
small-to-moderate $\tau$ values have 
also been observed and discussed 
in Refs.~\cite{fin14,els15, sch19}.
One might be tempted to conjecture that the few-body character of the scrambling operator $V$ in those examples invalidates a faithful modeling in terms of the random matrix ensemble introduced below Eq.~\eqref{eq:IIC:TimeEvoBack:A} in this regime.
However, we observed no such deviations in the setting of Fig.~\ref{fig2}, involving the same $H$ and $V$ as in Fig.~\ref{fig:IIC:SpinES}, but different $A$ and $\rhoTarget$.
Hence the few-body form of $V$ alone does not explain the deviations.
Since such small-$\tau$ deviations are also absent
in a ``direct'' implementation of the perturbation ensemble from Sec.~\ref{s4a}, which we demonstrate by means of example in Appendix~\ref{app3}, they hint at (mild) correlations between the observable $A$ and the initial state $\rhoTarget$ on the one hand and the Hamiltonian $H$ and the scrambling operator 
$V$ on the other hand.
Crucially,
by further increasing
$\tau$ in the examples from 
Fig.~\ref{fig:IIC:SpinES} or Appendix~\ref{app3}, we found, 
as predicted above, that the 
agreement between numerics and 
theory remains unchanged.
In particular, the persistence of the
echo peak signal is maintained.

\section{Imperfect reversal and combined imperfections}
\label{s5}
As pointed out in Sec. \ref{s3}, a second basic type of experimental
inaccuracy in the considered echo dynamics consists in an 
imperfect implementation of the time-reversed Hamiltonian \cite{sch16}.
Such an ``imperfect reversal'' scenario is particularly relevant with 
regard to experimental applications because (a) it is usually 
only feasible to invert the dominant part of the respective Hamiltonian, 
and (b) the accuracy to carry out the elaborate protocols 
necessary to do so is subject to experimental constraints 
\cite{sch69, rhi70, rhi71, zha92, kim92, haf96, lev98, usa98}. 
In fact, inverting only parts of the Hamiltonian can even be desirable 
as the degree of echo-signal attenuations can reveal information about 
the ``imperfections'' and thus the structure of the probed material,
the paradigmatic example being magnetic resonance imaging (MRI).

To separate the influences of such an imperfect reversal procedure 
from the imperfect preparation scenario discussed in Sec.~\ref{s4},
we assume for now that the return state is set up perfectly for the 
time-reversed evolution.
However, instead of the clean backward Hamiltonian $-H$, we 
consider a perturbed $H' = -H + \epsilon W$, leading to a total 
echo process of the form
\begin{equation}
\label{eq:ITR:Protocol}
	\rhoTarget \xrightarrow[H]{\;\;\;\;\tau\;\;\;\;} \rhoRev \xrightarrow[-H + \epsilon W]{\;\;\;\;\tau\;\;\;\;} \rhoTargetPert \,.
\end{equation}
Formally, this protocol may be regarded as a generalization of 
the Loschmidt echo \cite{per84, gor06, dor13}, since the latter 
is recovered (as a function of $\tau$) for the special
choice $A = \rhoTarget$ 
(whose experimental realization in a many-body system
may be difficult in general \cite{sch16,sch18,sch19}
but not impossible in specific examples \cite{dor13}).
Combinations of the two scenarios from~(\ref{eq:IIC:Protocol}) 
and~(\ref{eq:ITR:Protocol}) will be addressed at the end of this 
section.

Similarly as in the previous section, 
we have in mind a given system with fixed Hamiltonian $H$ 
whose dynamics is subject to some (partially)
uncontrolled or unknown experimental 
imperfection, described by $W$ in (\ref{eq:ITR:Protocol}).
Accordingly, this lack of control is again modeled by choosing 
an appropriate ensemble of random perturbations $W$, whose 
characteristics should nevertheless still emulate the key 
features of typical
imprecisions in many-body systems.
Notably, we therefore allow for a sparse distribution of the matrix 
elements $\bra{m} W \ket{n}$ and a dependence of their amplitude 
on the energy difference $E_m - E_n$ of the involved levels 
\cite{gen12, beu15, kon15, bor16, jan19}.
In other words, the random matrices $\bra{m} W \ket{n}$ 
may (but need not) exhibit a so-called sparse and/or banded 
structure~\cite{fyo96}.

The just-described setup was laid out and investigated in detail in 
Ref.~\cite{dab20}, so that we only summarize the key findings here.
Moreover, the conceptual approach is also very similar to the 
one from the previous Sec.~\ref{s4}:
The first step consists in averaging the echo dynamics 
from the protocol~\eqref{eq:ITR:Protocol} over the 
considered ensemble of perturbations $W$.
Denoting this average by $\avv{\,\cdots}$, 
too, the so-obtained average suppression 
of the echo signal then takes the form
\begin{equation}
\label{eq:ITR:AvgEcho}
	\frac{ \avv{ \dA(\tau + t) } }{ \dA(\tau - t) }
		= \e^{-2 \alpha t \epsilon^2 }
		\qquad
		(0 \leq t \leq \tau) \,,
\end{equation}
where $\alpha = \pi \bar\sigma_w^2 D_0$ is a constant involving 
the variance $\bar\sigma_w^2$ of the matrix elements $\bra{m} W \ket{n}$ 
for small $E_m - E_n$. Furthermore, $D_0$ is the density 
of states of $H$ 
in the vicinity of the system energy $E$ 
(see also Sec.~\ref{s4a}).

Similarly as for the result from Sec.~\ref{s4b}, this average~(\ref{eq:ITR:AvgEcho}) gains predictive power in systems 
with many degrees of freedom because the corresponding
variance is found to satisfy the bound \cite{dab20}
\begin{equation}
\label{eq:ITR:Variance}
	\avv{ \dA(\tau \!+\! t)^2 } - \avv{ \dA(\tau \!+\! t) }^{\,2}
		\leq \frac{c \,\da^2}{N_{w}} \,,
\end{equation}
where $c$ is a constant of the order of $10^3$ or smaller,
and $N_{w} := 2\alpha \epsilon^2 D_0$ quantifies how many 
unperturbed energy levels the perturbation $W$ couples 
on average.
Crucially, this latter number usually scales exponentially 
with the system's degrees of freedom \cite{dabYY}, so 
that~\eqref{eq:ITR:Variance} indeed becomes exceedingly 
small for typical many-body systems.

Altogether, we thus can conclude that in a system with sufficiently 
many degrees of freedom, nearly all imperfections 
$W$ in the time-reversed Hamiltonian which can be 
faithfully modeled by any of the above specified random 
matrix ensembles lead to an attenuation of the echo 
signal of the form \cite{dab20}
\begin{equation}
\label{eq:ITR:TypEcho}
	\frac{ \dA(\tau + t) }{ \dA(\tau - t) }
		= \e^{-2 \alpha t \epsilon^2 }
		\qquad
		(0 \leq t \leq \tau) \,.
\end{equation}
Contrary to the imperfect initial conditions [see Eq.~\eqref{eq:IIC:TypEcho}], 
the intensity of the echo signal at $t \approx \tau$ now decreases as the reversal 
time $\tau$ grows.
Essentially, this may be understood as a consequence of the fact that
the system is continuously exposed to the perturbation $W$
during the entire backward evolution of duration $\tau$.

Finally, it is conceivable and in fact practically inevitable 
that both types of imperfections discussed here 
actually contribute to the deterioration of an echo signal.
Analogously to~(\ref{eq:IIC:Protocol}) and~(\ref{eq:ITR:Protocol}), 
such a scenario may be symbolically indicated as
\begin{equation}
\label{x4}
	\rhoTarget \xrightarrow[H]{\;\;\;\;\tau\;\;\;\;} \rhoRev 
	\xrightarrow[V]{\;\;\;\;\delta\;\;\;\;} \rhoRevPert \xrightarrow[-H + \epsilon W]{\;\;\;\;\tau\;\;\;\;} \rhoTargetPert 
	\ .
\end{equation}
Provided that those two types of imperfections have 
``independent'' origins
(corresponding to statistically uncorrelated randomization 
effects of $V$ and $W$), 
the results from Secs.~\ref{s4} and~\ref{s5} can be 
readily combined by multiplying the right-hand 
sides of Eqs.~\eqref{eq:IIC:TypEcho}
and~\eqref{eq:ITR:TypEcho} to obtain a 
prediction for the total effect of the form
\begin{equation}
\label{eq:Both:TypEcho}
	\frac{ \dA(\tau + \delta + t) }{ \dA(\tau - t) }
		= \lvert \dosVft(\delta) \rvert^2 \, \e^{-2 \alpha t \epsilon^2} \,.
\end{equation}
While both types of imperfections thus generically cause a mitigation of the 
echo signal, only the persistent exposure to perturbations of the imperfect reversal 
type from this section entails a stronger decay with increasing reversal 
time $\tau$.
\\[0.3cm]
\section{Summary and Conclusions}
\label{s6}
At the focus of our present work is the echo 
dynamics of large but finite many-body quantum systems
when weakly perturbed according to the so-called
imperfect preparation scheme from (\ref{eq:IIC:Protocol}):
An initial state $\rhoTarget$ 
evolves under the action
of a Hamiltonian $H$ during a time period $\tau$,
is then slightly perturbed 
by a scrambling 
operator $V$ for a short time $\delta$,
and is finally propagated by the inverted
Hamiltonian for another time period $\tau$.
The key question is:
How does such an imperfect 
echo dynamics 
manifest itself in the time-dependent 
expectation values of an observable $A$,
and, in particular, how do the deviations
from a perfectly unperturbed echo dynamics
depend on the propagation time $\tau$,
the perturbation time $\delta$, and 
on the details of the perturbation
operator $V$?
As our main result we predicted that 
most perturbations $V$, which exhibit some 
rather weak general properties, will satisfy 
in very good approximation the relation 
(\ref{eq:IIC:TypEcho}), connecting the expectation values
of $A$ from (\ref{eq:DevAFromEq}) with the Fourier-transformed 
density of states of $V$ from (\ref{eq:DOS:V}) and (\ref{eq:DOS:V:FT}).
Figs.~\ref{fig:IIC:SpinES} and~\ref{fig2} illustrate
the very good agreement of this prediction with numerically 
exact results without any free fit parameter.

The most remarkable feature of Eq.~(\ref{eq:IIC:TypEcho}) 
is that the right-hand side is independent of 
$t$ and $\tau$.
In particular, for $t=\tau$ 
the deviation between the initial and the final expectation values
in (\ref{x5})
is independent of the propagation time 
$\tau$ in (\ref{eq:IIC:Protocol}), see also 
Fig. \ref{fig:IIC:SpinES} for an explicit example.
In contrast to the classical case,
where echo signals usually attenuate 
with increasing $\tau$ \cite{wij12, wij13},
the quantum echo dynamics is thus found to be of 
a remarkably persistent character.
Some heuristic arguments in support of
this property have been provided in 
Sec.~\ref{s4d} as well as in 
Ref.~\cite{sch19},  but for a fully
satisfying explanation, one apparently 
cannot circumvent the non-trivial
calculational details from Sec. \ref{s4b} 
and Appendix~\ref{app2}.

For the sake of completeness, we also
summarized in Sec. \ref{s5} our previously
obtained results from \cite{dab20}
for the so-called imperfect reversal 
scenario (\ref{eq:ITR:Protocol}), 
and we provided the pertinent generalization 
for cases which simultaneously exhibit an
imperfect preparation and an 
imperfect reversal, cf. (\ref{x4}).

As discussed at the end of Sec.~\ref{s4d},
our present work is complementary to
the earlier studies from Refs. 
\cite{fin14,els15}, where the main
focus was on small-to-moderate 
propagation times $\tau$ in (\ref{eq:IIC:Protocol}).
Moreover, our present approach is
also complementary to the one adopted 
in Refs.~\cite{sch16,sch18,sch19},
where the main focus was on the
echo dynamics in the thermodynamic
limit,
probing to what extent classical chaos 
indicators reemerge there \cite{sch16, sch18},
and 
how this limit is related
to the semiclassical limit 
$\hbar\to 0$ \cite{sch19}.
In passing, we note that also
Refs.~\cite{fin14,els15} had 
mainly in mind this limiting
case in the theoretical discussions,
while the numerical examples
were actually closer in spirit to our 
present setup.
Furthermore, it may be recalled that the 
concept of chaos in classical systems usually
does not involve the thermodynamic limit, 
but rather is mostly considered in systems 
of finite and possibly even low dimensionality.

While our findings here suggest that one 
signature of classical chaos 
in finite-dimensional systems,
namely sensitive dependence on initial 
conditions, may not have a corresponding 
quantum analog, we
finally mention that
there are various
well-established or recently proposed
alternative indicators of chaos (or nonintegrability)
in few- and many-body quantum systems,
such as level statistics \cite{bor16}, 
eigenstate thermalization \cite{dal16},
or out-of-time-order correlators 
\cite{mal16,swi18}, 
to name but a few.
Since we are not aware of any tangible
mathematical or physical connection with 
our present approach, we desist from a
more detailed compilation regarding the
agreement or disagreement with all those
alternative indicators for the 
``chaoticity'' of any given specific 
model.

\begin{acknowledgments}
We thank Boris Fine and Stefan Kehrein 
for inspiring discussions.
This work was supported by the 
Deutsche Forschungsgemeinschaft (DFG)
within the Research Unit FOR 2692
under Grant No. 397303734
and by the Paderborn Center for Parallel 
Computing (PC$^2$) within the Project 
HPC-PRF-UBI2.
\end{acknowledgments}

\appendix
\section{Derivation of Eq.~\eqref{eq:IIC:Variance}}
\label{app2}
In this appendix, we derive the bound~\eqref{eq:IIC:Variance} on the fluctuations of $\dA(\tau + \delta + t)$ under imperfect preparation of the return state as studied in Sec.~\ref{s4}.
We first observe that
\begin{equation}
\begin{aligned}
	& \avvu{ \dA(\tau \!+\! \delta \!+\! t)^2 } - \avvu{ \dA(\tau \!+\! \delta \!+\! t) }^{\,2}
	\\ & \;\;
		= \avvu{ ( \< A \>_{\!\rhoBackward(t)} )^2 } - \avvu{ \< A \>_{\!\rhoBackward(t)} }^{\,2} \,.
\end{aligned}
\end{equation}
Combining Eqs.~\eqref{eq:IIC:TimeEvoBack:rho} and~\eqref{eq:IIC:TimeEvoBack:A},
we can write
\begin{equation}
\label{eq:app:IIC:TimeEvo}
\begin{aligned}
	\< A \>_{\!\rhoBackward(t)}
		&\!=\! \sum_{\mu_1 \mu_2} \sum_{k_1, k_2} \sum_{l_1, l_2}
			\e^{\I (E_{l_1} - E_{k_2}) t} \e^{\I (E^V_{\mu_2} - E^V_{\mu_1}) \delta} \e^{\I (E_{l_2} - E_{k1}) \tau} \\
		& \qquad \times \bra{k_1} \rhoTarget \ket{l_2} \bra{k_2} A \ket{l_1} \,
			U_{\mu_1 k_1} U_{\mu_2 k_2} U_{\mu_1 l_1}^* U_{\mu_2 l_2}^*
\end{aligned}
\end{equation}
for the time-dependent expectation values during the backward evolution.
As derived in the main text, cf.~Eq.~\eqref{eq:IIC:AvgTimeEvo:rho2}, the average over all perturbations $V$ of this expression is given by
\begin{equation}
\label{eq:app:IIC:AvgTimeEvo}
\begin{aligned}
	\avvu{ \< A \>_{\!\rhoBackward(t)} }
		&= \left( \< A \>_{\!\rhoForward(\tau - t)} - \< A \>_{\rhoMC} \right) \lvert \dosVft(\delta) \rvert^2 \\
		& \qquad	+ \< A \>_{\rhoMC} - \frac{1}{N} \< A \>_{\!\rhoForward(\tau - t)} \,,
\end{aligned}
\end{equation}
from which we can read off $\avu{ \< A \>_{\!\rhoBackward(t)} }^{\,2}$ straightforwardly.
Squaring~\eqref{eq:app:IIC:TimeEvo} leads to
\begin{widetext}
\begin{equation}
\label{eq:app:IIC:AvgTimeEvoSquared}
\begin{aligned}
( \< A \>_{\!\rhoBackward(t)} )^2
		&= \sum_{\mu_1 \cdots \mu_4} \sum_{k_1 \cdots k_4} \sum_{l_1 \cdots l_4}
			\e^{\I (E_{l_1} - E_{k_2} + E_{l_3} - E_{k_4}) t} \e^{\I (E^V_{\mu_2} - E^V_{\mu_1} + E^V_{\mu_4} - E^V_{\mu_3}) \delta} \e^{\I (E_{l_2} - E_{k1} + E_{l_4} - E_{k3}) \tau} \\
		& \qquad \times \bra{k_1} \rhoTarget \ket{l_2} \bra{k_3} \rhoTarget \ket{l_4} \bra{k_2} A \ket{l_1} \bra{k_4} A \ket{l_3} \,
			U_{\mu_1 k_1} U_{\mu_2 k_2} U_{\mu_3 k_3} U_{\mu_4 k_4} U_{\mu_1 l_1}^* U_{\mu_2 l_2}^* U_{\mu_3 l_3}^* U_{\mu_4 l_4}^* \,.
\end{aligned}
\end{equation}
\end{widetext}
Computing the average thus amounts to averaging over the eight factors of $U_{\mu k}$.
Following Ref.~\cite{bro96}, this average is given by a sum over all possible combinations of pairing up the first and second indices of $U$ and $U^*$ factors, i.e.,
\begin{equation}
\label{eq:app:ICC:AvgU8}
\begin{aligned}
	&\avu{ U_{\mu_1 k_1} U_{\mu_2 k_2} U_{\mu_3 k_3} U_{\mu_4 k_4} U_{\mu_1 l_1}^* U_{\mu_2 l_2}^* U_{\mu_3 l_3}^* U_{\mu_4 l_4}^* } \\
		&\;\; = \sum_{P, P' \in \mathrm{Sym}(4)} \!\!\!\!\!\! v_{P,P'} \prod_{j=1}^4 \delta_{\mu_j \mu_{P(j)}} \delta_{k_j l_{P'(j)}} \,. 
\end{aligned}
\end{equation}
Here $\mathrm{Sym}(4)$ denotes the symmetric group of order $4$, i.e., the set of all permutations of $(1, 2, 3, 4)$, so that the sum in~\eqref{eq:app:ICC:AvgU8} consists of $(4!)^2 = 576$ terms.
The symmetry factors $v_{P, P'}$ depend only on the cyclic 
structure of $P^{-1} P'$.
There are five different combinations of cycle lengths in $\mathrm{Sym}(4)$, and the leading order as $N \gg 1$ of the corresponding factors for CUE or COE matrices is \cite{bro96}
\begin{equation}
\label{eq:app:IIC:SymmetryFactors4}
\begin{aligned}
	& v_{1,1,1,1} \simeq N^{-4} \,,\quad
	  v_{2,1,1} \simeq -N^{-5} \,, \\
	& v_{2,2} \simeq N^{-6} \,, \quad
	  v_{3,1} \simeq 2 N^{-6} \,, \quad
	  v_{4} \simeq -5 N^{-7} \,.
\end{aligned}
\end{equation}
As announced in the main text, we restrict to this leading order analysis here.
Substituting~\eqref{eq:app:ICC:AvgU8} into~\eqref{eq:app:IIC:AvgTimeEvoSquared}, 
the sums over $\mu_j$ and $k_j, l_j$ factorize, so that we can analyze them individually.
The factors involving $\mu_j$ have the general form
\begin{equation}
\label{eq:app:IIC:muTerms}
	F_P := \sum_{\mu_1 \cdots \mu_4} \e^{\I (E^V_{\mu_2} - E^V_{\mu_1} + E^V_{\mu_4} - E^V_{\mu_3}) \delta} \prod_{j=1}^4 \delta_{\mu_j \mu_{P(j)}} 
\end{equation}
with $P \in \mathrm{Sym}(4)$.
The order of these terms in $N$ is again determined by the cyclic structure of the relevant permutation $P$.
The different contributions are summarized in Tab.~\ref{tab:app:IIC:muTerms}.
\begin{table}
\caption{Contributions to the average of Eq.~\eqref{eq:app:ICC:AvgU8} from the 
sums over $\mu_j$ for the different permutations $P \in \mathrm{Sym}(4)$, 
cf.~Eq.~\eqref{eq:app:IIC:muTerms}.
}
\label{tab:app:IIC:muTerms}
\begin{tabularx}{\linewidth}{l|X|l}
\hline \hline
cycles & $P$ & $F_P$ \\
\hline
$1,1,1,1$ & $(1)(2)(3)(4)$ & $N^4 \, \lvert \dosVft(\delta) \rvert^4$ \\
$2,1,1$   & $(1 \; 2)(3)(4)$, $(1 \; 4)(2)(3)$, $(1)(2 \; 3)(4)$, $(1)(2)(3 \; 4)$  & $N^3 \,  \lvert \dosVft(\delta) \rvert^2$ \\
          & $(1 \; 3)(2)(4)$ & $N^3 \, \dosVft(\delta)^2 \, \dosVft^*(2\delta)$ \\
          & $(1)(2 \; 4)(3)$ & $N^3 \, \dosVft(2 \delta) \, \dosVft^*(\delta)^2$ \\
$2,2$     & $(1 \; 2)(3 \; 4)$, $(1 \; 4)(2 \; 3)$ & $N^2$ \\
          & $(1 \; 3)(2 \; 4)$ & $N^2 \, \lvert \dosVft(2\delta)\rvert^2$ \\
$3,1$     & $(1 \; 2 \; 3)(4)$, $(1 \; 3 \; 2)(4)$, $(1 \; 2 \; 4)(3)$, $(1 \; 4 \; 2)(3)$, $(1 \; 3 \; 4)(2)$, $(1 \; 4 \; 3)(2)$, $(1)(2 \; 3 \; 4)$, $(1)(2 \; 4 \; 3)$ & $N^2 \, \lvert \dosVft(\delta) \rvert^2$ \\
$4$       & $(1 \; 2 \; 3 \; 4)$, $(1 \; 2 \; 4 \; 3)$, $(1 \; 3 \; 2 \; 4)$, $(1 \; 3 \; 4 \; 2)$, $(1 \; 4 \; 2 \; 3)$, $(1 \; 4 \; 3 \; 2)$ & $N$ \\
\hline \hline
\end{tabularx}
\end{table}
The general form of the factors with $k_j, l_j$ for a permutation $P' \in \mathrm{Sym}(4)$ is
\begin{widetext}
\begin{equation}
\label{eq:app:IIC:klTerms}
	G_{P'} := \sum_{k_1 \cdots k_4} \sum_{l_1 \cdots l_4}
			\e^{\I (E_{l_1} - E_{k_2} + E_{l_3} - E_{k_4}) t}
			\e^{\I (E_{l_2} - E_{k1} + E_{l_4} - E_{k3}) \tau} 
			\bra{k_1} \rhoTarget \ket{l_2} \bra{k_3} \rhoTarget \ket{l_4} \bra{k_2} A \ket{l_1} \bra{k_4} A \ket{l_3}
			\prod_{j=1}^4 \delta_{k_j l_{P'(j)}}
\end{equation}
\end{widetext}
These reduce to the 10 different combinations of state and observable summarized in Tab.~\ref{tab:app:IIC:klTerms},
of which there are two of order $N^2$, four of order $N$, and five of order $1$.
\begin{table}
\caption{Contributions to the average of Eq.~\eqref{eq:app:ICC:AvgU8} 
from the sums over $k_j, l_j$ for the different permutations 
$P' \in \mathrm{Sym}(4)$, cf.~Eq.~\eqref{eq:app:IIC:klTerms}.
}
\label{tab:app:IIC:klTerms}
\begin{tabularx}{\linewidth}{X|l}
\hline \hline
$P'$ & $G_{P'}$ \\
\hline
$(1)(2)(3)(4)$, $(1 \; 3)(2 \; 4)$ & $(\< A \>_{\!\rhoForward(\tau - t)} )^2$ \\
$(1 \; 2)(3)(4)$, $(1)(2)(3 \; 4)$, $(1 \; 2 \; 4 \; 3)$, $(1 \; 3 \; 4 \; 2)$ & $N\! \< A \>_{\!\rhoMC} \< A \>_{\!\rhoForward(\tau - t)}$ \\
$(1 \; 2)(3 \; 4)$ & $N^2 (\< A \>_{\!\rhoMC})^2$ \\
$(1)(2 \; 3)(4)$, $(1 \; 4)(2)(3)$, $(1 \; 3 \; 2 \; 4)$, $(1 \; 4 \; 2 \; 3)$ & $\tr[ A^2 \rhoForward(\tau - t)^2 ]$ \\
$(1 \; 2 \; 3)(4)$, $(1)(2 \; 3 \; 4)$, $(1 \; 2 \; 4)(3)$, $(1 \; 3 \; 4)(2)$ & $N\! \< A \>_{\!\rhoMC} \!\tr[ A \rhoForward(\tau-t)^2 ]$ \\
$(1 \; 3 \; 2)(4)$, $(1 \; 4 \; 2)(3)$, $(1 \; 4 \; 3)(2)$,  $(1)(2 \; 4 \; 3)$ & $\< A^2 \>_{\!\rhoForward(\tau-t)}$ \\
$(1 \; 3)(2)(4)$, $(1)(2 \; 4)(3)$ & $\tr\{ [ \rhoForward(\tau-t) A ]^2 \}$ \\
$(1 \; 2 \; 3 \; 4)$ & $N^2\! \< A \>_{\!\rhoMC} \! \tr[ \rhoTarget^2 ]$ \\
$(1 \; 4 \; 3 \; 2)$ & $N\! \< A^2 \>_{\!\rhoMC}$ \\
$(1 \; 4)(2 \; 3)$ & $N\! \< A^2 \>_{\!\rhoMC} \! \tr[ \rhoTarget^2 ]$ \\
\hline \hline
\end{tabularx}
\end{table}

Combining the factors $v_{P,P'}$ from~\eqref{eq:app:IIC:SymmetryFactors4}, $F_P$ from~\eqref{eq:app:IIC:muTerms} and Tab.~\ref{tab:app:IIC:muTerms}, and $G_{P'}$ from~\eqref{eq:app:IIC:klTerms} and Tab.~\ref{tab:app:IIC:klTerms},
there are nine terms of order $1$ in the average of~\eqref{eq:app:IIC:AvgTimeEvoSquared}.
These cancel exactly against the nine terms of order $1$ obtained by squaring~\eqref{eq:app:IIC:AvgTimeEvo}, so that the variance $\avu{ ( \langle A \rangle_{\!\rhoBackward(t)} )^2 } - \avu{ \langle A \rangle_{\!\rhoBackward(t)} }^{\,2}$ vanishes to order $1$.

At order $N^{-1}$, there are $42$ terms in the average of~\eqref{eq:app:IIC:AvgTimeEvoSquared} and six terms in the square of~\eqref{eq:app:IIC:AvgTimeEvo}, giving
\begin{widetext}
\begin{equation}
\begin{aligned}
	& \avvu{ \left( \< A \>_{\!\rhoBackward(t)} \right)^2 } - \avvu{ \< A \>_{\!\rhoBackward(t)} }^{\,2}
		= \frac{1}{N} \left\{
				2 \< A \>_{\!\rhoForward(\tau - t)} \< A \>_{\!\rhoMC} - ( \< A \>_{\!\rhoMC} )^2 \tr(\rhoTarget^2) 
				\right. \\ & \left. \quad			
				+ 2 \lvert \dosVft(\delta) \rvert^2 \left[ ( \< A \>_{\!\rhoForward(\tau - t)} )^2 - \< A \>_{\!\rhoForward(\tau - t)} \< A \>_{\!\rhoMC} + \tr[ A^2 \rhoForward(\tau - t)^2 ] - 2 \tr[ A \rhoForward(\tau - t)^2 ] \< A \>_{\!\rhoMC} + 2 (\< A \>_{\!\rhoMC})^2 \tr(\rhoTarget^2) \right]
				\right. \\ & \left. \quad
				+ \left[ \dosVft(2\delta) \dosVft(\delta)^{*2} + \dosVft(\delta)^2 \dosVft(2 \delta)^* \right] \left[ (\< A \>_{\!\rhoMC})^2 \tr(\rhoTarget^2) - 2 \tr[ A \rhoForward(\tau - t)^2 ] + \tr\{ [ \rhoForward(\tau-t) A ]^2 \} \right]
				\right. \\ & \left. \quad
				- \lvert \dosVft(\delta) \rvert^4 \left[ 2 \tr[ A^2 \rhoForward(\tau - t)^2 ] - 8 \tr[ A \rhoForward(\tau - t)^2 ] \< A \>_{\!\rhoMC} + 2 \tr\{ [ \rhoForward(\tau-t) A ]^2 \} + 5 (\< A \>_{\!\rhoMC})^2 \tr(\rhoTarget^2) \right]
			\right\}
		+ \mathcal{O}\!\left(\frac{1}{N^2}\right)
\end{aligned}
\end{equation}
\end{widetext}
Using the triangle inequality, the inequality $\lvert \dosVft(\delta) \rvert \leq 1$, 
and bounding the absolute value of all the different traces over 
factors of $A$ and $\rho$ by $\opnorm{A}^2$ (operator norm),
we then arrive at the estimate
\begin{equation}
\label{eq:a2}
	\avvu{ ( \< A \>_{\!\rhoBackward(t)} )^2 } - \avvu{ \< A \>_{\!\rhoBackward(t)} }^{\,2}
		\leq \frac{42 \, \opnorm{A}^2}{N} + \mathcal{O}\!\left(\frac{1}{N^2}\right) 
		\ .
\end{equation}

As in the main text, we tacitly focus on cases
where $A$ models an observable with a finite 
measurement range $\da$ (difference between 
largest and smallest eigenvalues of $A$,
see below (\ref{eq:IIC:Variance}))
and thus with finite $\opnorm{A}$.
Without loss of generality, we can and will
assume (possibly after adding an irrelevant 
constant to $A$) that the largest and smallest 
eigenvalues of $A$ are equal in modulus and 
of opposite sign, implying that
$\opnorm{A}=\da/2$. Together with 
(\ref{eq:a2}), this finally yields 
Eq.~\eqref{eq:IIC:Variance} from the main text.


\section{Further numerical examples}
\label{app3}
\begin{figure}
\includegraphics[scale=1]{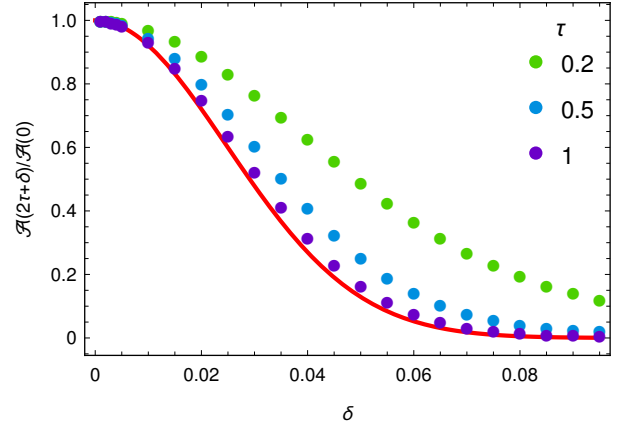}
\caption{
Same as in the insets of Fig.~\ref{fig:IIC:SpinES},
but now for three rather small reversal times
$\tau=0.2,\, 0.5,\, 1$.
In other words, the ratio $\mc A(2\tau + \delta) / \mc A(0)$ 
of echo and initial peak heights is depicted as a function 
of the scrambling time $\delta$
for the staggered magnetization~(\ref{eq:IIC:Example:A})
of a spin-$\frac{1}{2}$ XXX chain~(\ref{eq:IIC:Example:H})
with $L = 14$, normally distributed $J_{ij}^{\alpha\beta}$
in~\eqref{eq:IIC:Example:V}, ``imperfect preparation'' 
protocol~\eqref{eq:IIC:Protocol}, and N\'{e}el target 
state $\rhoTarget$.
The colored dots are the numerical solutions
for increasing $\tau$ from top to bottom, 
the solid red curve is the analytical prediction 
from~\eqref{eq:IIC:TypEcho}, \eqref{x}, and 
\eqref{eq:IIC:Example:DHat}.
}
\label{fig3}
\end{figure}

\begin{figure}
\includegraphics[scale=1]{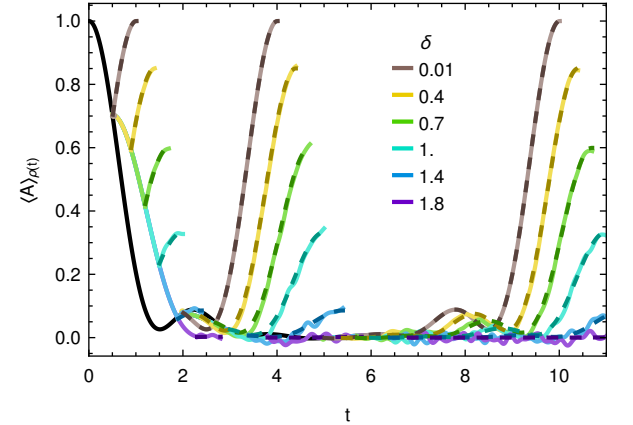}
\caption{
Time-dependent expectation value of a GOE observable $A$ for 
the system~\eqref{eq:RandEx:H}, starting from the state 
$\rhoTarget = \ket{\psi}\bra{\psi}$ 
from~\eqref{eq:RandEx:RhoTarget}.
Results are shown for various scrambling 
parameters $\delta$ (color-coded as indicated, with the peak height decreasing as $\delta$ increases) and reversal 
times $\tau = 0.5, 2, 5$.
Solid lines correspond to the numerics, dashed lines are 
the theoretical predictions for the backward evolution 
according to~\eqref{eq:IIC:TypEcho} 
with~\eqref{eq:RandEx:DOSVFT} and 
$\langle A \rangle_{\rhoMC} = 0$.
The echo peak height is entirely independent 
of $\tau$.
}
\label{fig4}
\end{figure}

In this appendix, we elaborate in more detail on the deviations
between our theoretical prediction~\eqref{eq:IIC:TypEcho} and 
the spin-model numerics from Sec.~\ref{s4c} that occur for 
$\tau \lesssim 1$ when starting from a N\'{e}el state and observing the staggered magnetization.
We emphasize that these deviations do not affect our main 
conclusion, namely the persistence of many-body quantum 
echoes at arbitrarily large times:
For $\tau \gtrsim 1$ and sufficiently small $\delta$, 
the echo peaks remain finite and become independent 
of $\tau$.
(For the examples of Figs.~\ref{fig:IIC:SpinES} 
and~\ref{fig2}, we checked reversal times up to $
\tau = 20$ and $\tau = 36$, respectively.)

We also emphasize that there are no such deviations for small $\tau$ in the setup from Fig.~\ref{fig2}, where the reference Hamiltonian $H$ and scrambling operator $V$ are identical to the ones from Fig.~\ref{fig:IIC:SpinES} [Eqs.~\eqref{eq:IIC:Example:H} and~\eqref{eq:IIC:Example:V}],
but the initial state and observable are superpositions of two energy eigenstates instead.

Coming back to the setup from Fig.~\ref{fig:IIC:SpinES},
we depict in Fig.~\ref{fig3}
the echo peak height in the 
spin model from Sec.~\ref{s4c} for reversal times 
$\tau = 0.2, 0.5, 1$ as a function of the scrambling 
parameter $\delta$ with the solid red line indicating the
theoretical prediction from Eq.~\eqref{eq:IIC:TypEcho}.
This figure should be contrasted with the insets of Fig.~\ref{fig:IIC:SpinES}.
For fixed $\delta$, the actual (numerical) echo signal 
is attenuated \emph{less} than our theory predicts, 
with the deviations becoming stronger as $\tau$ 
decreases.

One reason for this may be that the return state 
$\rhoRev$ for small $\tau$ violates the assumption 
[see below Eq.~\eqref{eq:IIC:LevelsInShell}] of 
being uniformly spread across the 
$V$ eigenvectors in the energy shell (see also 
the discussion at the end 
of Sec.~\ref{s4d}), i.e., the special structure 
of the N\'{e}el initial state is only gradually 
stirred up during the forward evolution.
A related problem can arise when the perturbation 
$V$ is correlated with $H$, $A$, and/or $\rhoTarget$, 
so that a reliable modeling in terms of the unbiased 
ensemble introduced in Sec.~\ref{s4a} is not 
guaranteed.
Such issues are expected to be most compromising 
when $V$ is applied on a state $\rhoRev$ that is 
still out of equilibrium (i.e., for small $\tau$) 
since this is the regime dominated by correlations.

Indeed, we observe that there are no small-$\tau$ 
deviations when one avoids all these correlations 
by choosing a sufficiently randomized system.
To this end, we consider a Hamiltonian
\begin{equation}
\label{eq:RandEx:H}
	H = \sum_n \frac{n}{D} \ket{n} \bra{n}
\end{equation}
with a fixed level spacing $1/D$ and corresponding
level density $D = 512$, in combination with
an observable $A$ which is randomly sampled 
from the Gaussian orthogonal ensemble (GOE).
The initial state $\rhoTarget = \ket{\psi}\bra{\psi}$ is generated by exploiting dynamical typicality \cite{bar09} from a Haar-distributed random Hilbert space vector $\ket{\phi}$ according to
\begin{equation}
\label{eq:RandEx:RhoTarget}
	\ket{\psi} \propto (1 + \gamma A) \ket{\phi}
\end{equation}
with $\gamma = 1$.
For the perturbation $V$, we draw a matrix at random from the Gaussian unitary ensemble (GUE), so that $\dosV(E)$ is a semicircle of radius $2$ and consequently
\begin{equation}
\label{eq:RandEx:DOSVFT}
	\dosVft(\delta) = \frac{J_1(2 \delta)}{\delta} \,,
\end{equation}
where $J_1(x)$ denotes the Bessel function of the first kind of order $1$.
Once again, we thus know all parameters entering the theoretical prediction~\eqref{eq:IIC:TypEcho} exactly and can compare the analytics with numerical simulations without any fit parameters.
Turning to Fig.~\ref{fig4} for this comparison, we find that the agreement is virtually flawless, irrespective of the magnitude of either $\tau$ or $\delta$.
In the absence of any correlations (apart from those between $A$ and $\rhoTarget$ necessary for nonequilibrium initial conditions), the small-$\tau$ deviations thus disappear, too.



\begin{thebibliography}{70}

\bibitem{ott94}
E. Ott,
{\em Chaos in Dynamical Systems},
Cambridge University Press, Cambridge (1994).

\bibitem{lic92}
A. J. Lichtenberg and M. A. Liebermann,
{\em Regular and Chaotic Dynamics},
Springer, New York (1992).

\bibitem{lak03}
M. Lakshmanan and S. Rajasekar,
{\em Nonlinear Dynamics: Integrability, Chaos and Patterns},
Springer, Berlin (2003).

\bibitem{dal16}
L. D'Alessio, Y. Kafri, A. Polkovnikov, and M. Rigol,
From Quantum Chaos and Eigenstate Thermalization
to Statistical Mechanics and Thermodynamics,
Adv. Phys.  {\bf 65}, 239 (2016).

\bibitem{gog16}
C. Gogolin and J. Eisert,
Equilibration, thermalization, and the emergence
of statistical mechanics in closed quantum systems,
Rep. Prog. Phys. {\bf 79}, 056001 (2016).

\bibitem{mor18}
T. Mori, T. N. Ikeda, E. Kaminishi, and M. Ueda,
Thermalization and prethermalization 
in isolated quantum systems: a theoretical overview,
J. Phys. B  {\bf 51}, 112001 (2018).

\bibitem{rei13}
P. Reimann and M. Evstigneev,
Quantum versus classical foundation of statistical mechanics 
under experimentally realistic conditions,
Phys. Rev. E {\bf 88}, 052114 (2013).

\bibitem{hah50}
E. L. Hahn, Spin Echoes,
Phys. Rev. {\bf 80}, 580 (1950).

\bibitem{wij12}
A. S. de Wijn, B. Hess, and B. V. Fine,
Largest Lyapunov exponents for lattices of interacting classical spins,
Phys. Rev. Lett. {\bf 109}, 034101 (2012).

\bibitem{wij13}
A. S. de Wijn, B. Hess, and B. V. Fine,
Lyapunov instabilities in lattices of interacting 
classical spins at infinite temperature,
J. Phys. A: Math Theor. {\bf 46}, 254012 (2013).

\bibitem{sch69}
H. Schneider and H. Schmiedel,
Negative time development of a nuclear spin system,
Phys. Lett. {\bf 30A}, 298 (1969).

\bibitem{rhi70}
W.-K. Rhim, A. Pines, and J. S. Waugh,
Violation of the spin-temperature hypothesis,
Phys. Rev. Lett. {\bf 25}, 218 (1970).

\bibitem{rhi71}
W.-K. Rhim, A. Pines, and J. S. Waugh,
Time-Reversal Experiments in Dipolar Coupled Spin Systems,
Phys. Rev. B {\bf 3}, 684 (1971).

\bibitem{zha92}
S. Zhang, B. H. Meier, and R. R. Ernst,
Polarization echoes in NMR,
Phys. Rev. Lett. {\bf 69}, 2149 (1992).

\bibitem{kim92}
R. Kimmich, J. Niess, and S. Hafner,
Quadrupolar magic echoes,
Chem. Phys. Lett. {\bf 190}, 503 (1992).

\bibitem{haf96}
S. Hafner, D. E. Demco, and R. Kimmich,
Magic echoes and NMR imaging of solids,
Solid State Nucl. Mag. Res. {\bf 6}, 275 (1996).

\bibitem{lev98}
P. R. Levstein, G. Usaj, and H. M. Pastawski,
Attenuation of polarization echoes in nuclear magnetic resonance: A study of the emergence of dynamical irreversibility in many-body quantum systems,
J. Chem. Phys. {\bf 108}, 2718 (1998).

\bibitem{usa98}
G. Usaj, H. M. Pastawski, and P. R. Levstein,
Gaussian to exponential crossover in the attenuation of polarization echoes in NMR,
Mol. Phys. {\bf 95}, 1229 (1998).

\bibitem{wid08}
A. Widera, S. Trotzky, P. Cheinet, S. F\"olling, F. Gerbier, I. Bloch, V. Gritsev, M. D. Lukin, and E. Demler,
Quantum Spin Dynamics of Mode-Squeezed Luttinger Liquids in Two-Component Atomic Gases,
Phys. Rev. Lett. {\bf 100}, 140401 (2008).

\bibitem{lin16}
D. Linnemann, H. Strobel, W. Muessel, J. Schulz, R. J. Lewis-Swan, K. V. Kheruntsyan, and M. K. Oberthaler,
Quantum-Enhanced Sensing Based on Time Reversal of Nonlinear Dynamics,
Phys. Rev. Lett. {\bf 117}, 013001 (2016).

\bibitem{gar17}
M. G\"arttner, J. G. Bohnet, A. Safavi-Naini, M. L. Wall, J. J. Bollinger, and A. M. Rey,
Measuring out-of-time-order correlations and multiple quantum spectra in a trapped ion quantum magnet,
Nat. Phys. {\bf 13}, 781 (2017).

\bibitem{nik20}
M. Niknam, L. F. Santos, and D. G. Cory,
Sensitivity of quantum information to environment perturbations measured with a nonlocal out-of-time-order correlation function,
Phys. Rev. Research {\bf 2}, 013200 (2020).

\bibitem{fin14}
B. V. Fine, T. A. Elsayed, C. M. Kropf, and A. S. de Wijn,
Absence of exponential sensitivity to small perturbations in nonintegrable systems of spin 1/2,
Phys. Rev. E {\bf 89}, 012923 (2014).

\bibitem{els15}
T. A. Elsayed and B. V. Fine, Sensitivity to 
small perturbations in systems of large quantum spins,
Phys. Scr. {\bf T165}, 014011 (2015)

\bibitem{sch16}
M. Schmitt and S. Kehrein,
Effective time reversal and echo dynamics in the transverse field Ising model,
EPL {\bf 115}, 50001 (2016).

\bibitem{sch18}
M. Schmitt and S. Kehrein,
Irreversible dynamics in quantum many-body systems,
Phys. Rev. B {\bf 98}, 180301 (2018).

\bibitem{sch19}
M. Schmitt, D. Sels, S. Kehrein, and A. Polkovnikov,
Semiclassical dynamics in the Sachdev-Ye-Kitaev model,
Phys. Rev. B {\bf 99}, 134301 (2019).

\bibitem{dab20}
L. Dabelow and P. Reimann,
Predicting imperfect echo dynamics in many-body quantum systems,
Z. Naturforsch. A {\bf 75}, 403 (2020).

\bibitem{rei08}
P. Reimann,
Foundations of statistical mechanics under experimentally realistic conditions,
Phys. Rev. Lett. {\bf 101}, 190403 (2008).

\bibitem{lin09}
N. Linden, S. Popescu, A. J. Short, and A. Winter,
Quantum mechanical evolution towards thermal equilibrium,
Phys. Rev. E {\bf 79}, 061103 (2009).

\bibitem{sho12}
A. J. Short, T. C. Farrelly,
Quantum equilibration in finite time,
New J. Phys. {\bf 14}, 013063 (2012).

\bibitem{rei12}
P. Reimann and M. Kastner, 
Equilibration of macroscopic quantum systems,
New J. Phys. {\bf 14}, 043020 (2012).

\bibitem{f1}
Here and in the following, 
the word ``scrambling'' is used in the naive 
sense of ``mixing'', ``stirring'', or ``perturbing'', 
not in the specific sense in which it
has recently been introduced
in the context of probing characteristic signatures 
of quantum chaos (sensitivity to small perturbations, 
delocalization of information) by
out-of-time-order correlators \cite{mal16,swi18}.

\bibitem{mal16}
J. Maldacena, S. H. Shenker, D. Stanford,
A bound on chaos,
J. High Energy Phys. {\bf 2016}, 106 (2016).

\bibitem{swi18}
B. Swingle,
Unscrambling the physics of out-of-time-order correlators,
Nat. Phys. {\bf 14}, 988 (2018).

\bibitem{lan70}
L. Landau and E. Lifshitz, 
{\em Statistical Physics}, Pergamon Oxford (1970).

\bibitem{rei19a}
P. Reimann,
Transportless equilibration in isolated many-body quantum systems, 
New J. Phys. {\bf 21}, 053014 (2019).

\bibitem{rei16}
P. Reimann,
Typical fast thermalization processes in closed many-body systems,
Nat. Commun. {\bf 7}, 10821 (2016).

\bibitem{haa10}
F. Haake,
{\em Quantum Signatures of Chaos},
Springer Berlin (2010).

\bibitem{bro96}
P. W. Brouwer and C. W. J. Beenakker,
Diagrammatic method of integration over the unitary group, with applications to quantum transport in mesoscopic systems,
J. Math. Phys. {\bf 37}, 4904 (1996).

\bibitem{tor14}
E. J. Torres-Herrera and L. F. Santos,
Quench dynamics of isolated many-body quantum systems,
Phys Rev. A {\bf 89}, 043620 (2014).

\bibitem{erd14}
L. Erd\H{o}s and D. Schr\"oder,
Phase transitions in the density of states of quantum spin glasses,
Math. Phys. Anal. Geom. {\bf 17}, 441 (2014).

\bibitem{per84}
A. Peres,
Stability of quantum motion in chaotic and regular systems,
Phys. Rev. A {\bf 30}, 1610 (1984).

\bibitem{gor06}
T. Gorin, T. Prosen, T. H. Seligman, and M. \v{Z}nidari\v{c},
Dynamics of Loschmidt echoes and fidelity decay,
Phys. Rep. {\bf 435}, 33 (2006).

\bibitem{dor13}
B. D\'{o}ra, F. Pollmann, J. Fort\'{a}gh, and G. Zar\'{a}nd,
Loschmidt Echo and the Many-Body Orthogonality Catastrophe in a Qubit-Coupled Luttinger Liquid,
Phys. Rev. Lett. {\bf 111}, 046402 (2013).

\bibitem{gen12}
S. Genway, A. F. Ho, and D. K. K. Lee,
Thermalization of local observables in small Hubbard lattices,
Phys. Rev. A  {\bf 86}, 023609 (2012).

\bibitem{beu15}
W. Beugeling, R. Moessner, and M. Haque,
Off-diagonal matrix elements of local operators in many-body quantum systems,
Phys. Rev. E {\bf 91}, 012144 (2015).

\bibitem{kon15}
N. P. Konstantinidis,
Thermalization away from integrability and the role of operator off-diagonal elements,
Phys. Rev. E {\bf 91}, 052111 (2015).

\bibitem{bor16}
F. Borgonovi, F. M. Izrailev, L. F. Santos, and V. G. Zelevinsky,
Quantum chaos and thermalization in isolated systems of 
interacting particles,
Phys. Rep.  {\bf 626}, 1 (2016).

\bibitem{jan19}
D. Jansen, J. Stolpp, L. Vidmar, and F. Heidrich-Meisner,
Eigenstate thermalization and quantum chaos in the Holstein polaron model,
Phys. Rev. B {\bf 99}, 155130 (2019).

\bibitem{fyo96}
Y. V. Fyodorov, O. A. Chubykalo, F. M. Izrailev, and G. Casati,
Wigner random banded matrices with sparse structure: local spectral density of states,
Phys. Rev. Lett.  {\bf 76}, 1603 (1996).

\bibitem{dabYY}
L. Dabelow and P. Reimann,
Relaxation Theory for Perturbed Many-Body Quantum Systems versus Numerics and Experiment,
Phys. Rev. Lett. {\bf 124}, 120602 (2020).

\bibitem{bar09}
C. Bartsch and J. Gemmer,
Dynamical Typicality of Quantum Expectation Values,
Phys Rev. Lett. {\bf 102}, 110403 (2009).

\end{thebibliography}
\end{document}